\documentclass[showpacs, aps, prb, unsortedaddress]{revtex4-1}

\usepackage{amssymb}
\usepackage{latexsym}
\usepackage{amsmath}
\usepackage{graphicx}
\begin{document}

\title{Fast phononic band-structure calculations through a GPU accelerated mixed-variational formulation}

\date{\today}
\author{Ankit Srivastava}
\thanks{Corresponding Author}
\email{ankit.srivastava@iit.edu}

\affiliation{Department of Mechanical, Materials, and Aerospace Engineering
Illinois Institute of Technology, Chicago, IL, 60616
USA}
\pacs{*43.20.Gp, *43.20.Jr, 62.20.D-}

\begin{abstract}
In this paper we present a Graphical Processing Unit (gpu) accelerated mixed variational formulation for fast phononic band-structure calculation of arbitrarily complex unit cells and report speed gains of a hundred fold over unoptimized serial cpu computations. To the author's knowledge this is the first application of gpu computing to a non-FE/FDTD bandstructure algorithm. The formulation is presented in a form which is  applicable to 1-, 2-, and 3-D cases. However, in this paper we concentrate on optimizing the formulation within the paradigm of gpu computing, presenting results for 2-D unit cells. The mixed-variational formulation has been proven to show faster convergence than variational principles based purely on the displacement field. Additionally the integral nature of the method allows for its application to unit cells of arbitrary complexity. However, the application of this formulation was limited to fairly simple unit cells until recently when its scope was extended to deal with all Bravais lattices and complex micro-structures. In this paper we describe the application of the formulation within the framework of gpu computing with a long term view towards highly efficient and massively distributed band-structure algorithms suitable for tackling optimization and inverse problems. We report that the current formulation becomes I/O bound as opposed to compute bound indicating the potential for yet faster computations through optimized data transfer between the host (cpu) and the device (gpu). We report that the accurate band-structure evaluation over the boundary of the Irreducible Brillouin Zone (IBZ) for the first 18 phononic branches of a complex 2-D unit cell (with 1132 different phases) can be achieved in less than 20 seconds on a regular desktop. For a simpler unit cell, the first few phononic branches are calculated in less than 2.5 seconds on the same system. The scheme presented in this paper, therefore, represents considerable savings in the computational costs of phononic (and photonic, through extensions to the current method) band-structures. We discuss ways by which the computations can be further accelerated and comment upon the potential use of the fast algorithm presented in this paper to the areas of band-structure optimization and the inverse problem of dynamic homogenization.
\end{abstract}

\maketitle

\section{Introduction}

There has been significant recent interest in achieving exotic dynamic response through the careful microstructural design of composites. The periodic modulation of stress waves results in very rich wave-physics and the potential for novel applications \cite{cervera2001refractive,yang2002ultrasound,khelif2003trapping,reed2003reversed,yang2004focusing,gorishnyy2005hypersonic,mohammadi2008evidence,sukhovich2008negative,lin2009gradient}. The study of the effects and the applications of such periodic modulations falls within the area of phononics \cite{martinezsala1995sound}. These applications depend upon the ability of calculating the phononic band-structure of unit cells. Recent years have seen extensive developments in the numerical and analytical schemes required for such calculations (See \cite{hussein2009reduced} for a list of references).

The research community is now at a point where in addition to being able to calculate phononic/photonic band-structures, it is desirable to have the capability of executing such computations highly efficiently. The areas which stand to benefit from such fast computations include phononic/photonic optimization and inverse problems in effective properties (See \cite{torquato2010optimal}). In photonics the computational requirement is the search for the eigenvalues of the Maxwell's equations and in phononics the requirement is the analogous search for the Navier equations. By their very nature, these searches, especially in 2-, and 3-D, are highly compute intensive tasks requiring significant computational and time resources. As such any optimization procedure for photonic/phononic crystals is, therefore, hard to implement. Some recent advances have been made in the areas of photonic \cite{sigmund2003systematic,rupp2007design,diaz2005design,halkjaer2006maximizing,sigmund2008geometric} and phononic \cite{bilal2011ultrawide} band-structure optimization. However, a distributed and efficient computational platform for fast band-structure calculations has not yet been developed. In this paper we present the mixed-formulation for phononic band-structure computation within the paradigm of gpu-computing. The formulation \cite{nemat1972harmonic} itself has been proven to show fast convergence \cite{babuska1978numerical} and has recently been extended to deal with arbitrary complexities in the unit cell \cite{srivastava2013mixed}. It, therefore, is an ideal candidate for fast-band-structure computations on distributed platforms. Graphical processing units, which are normally used to render graphics on laptops and desktops, have also emerged as a cheap and efficient alternative to massive clusters for performing computational tasks which require a large number of computational processes to be run in parallel. Their suitability for parallel processing algorithms has been exploited in fields as varied as protein folding , molecular dynamics, calculating electrostatic potential maps \cite{elsen2007n,stone2007accelerating,owens2008gpu}, and weather prediction\cite{michalakes2008gpu}. However, the use of gpus in the area of band-structure calculation (phononic or photonic) has been limited to few applications to the finite difference time domain method \cite{humphrey2006high}. In this paper we report that the gpu-accelerated version of the mixed-formulation shows speed-ups of more than 2 orders of magnitude compared to the serial implementation (unoptimized) of the algorithm. To the author's knowledge, this represents the first time that gpus have been used to accelerate a non FE/FDTD band-structure formulation. We report sub-20 second time (on a regular desktop PC) for the complete phononic characterization over the first 18 branches of a 2-D hexagonal unit cell made up of more than thousand different phases. For a simpler unit cell and fewer branches, we report computation time of less than 2.5 second for the complete phononic characterization over the IBZ.

These results represent considerable savings in computation times and resources and indicate that the algorithm can provide very fast band-structure calculation capabilities for more complex 3-D systems through efficient memory management techniques. Extensions to the gpu-accelerated algorithm for fast dynamic effective property calculations \cite{nemat2011overall,srivastava2012overall,shuvalov2011effective,willis2011effective,willis2012construction} are natural with the aim being its application to topology optimization problems for dynamic effective properties instead of static effective properties \cite{guest2006optimizing,guest2007design}.

\section{Statement of the problem}

\begin{figure}[htp]
\centering
\includegraphics[scale=.6]{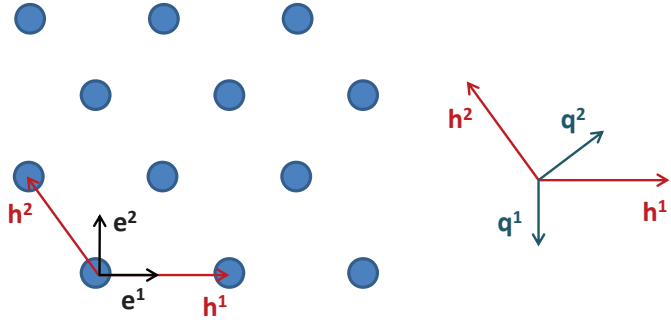}
\caption{Schematic of a 2-dimensional periodic composite. The unit cell vectors ($\mathbf{h}^1,\mathbf{h}^2$), reciprocal basis vectors ($\mathbf{q}^1,\mathbf{q}^2$), and the orthogonal vectors ($\mathbf{e}^1,\mathbf{e}^2$) are shown.}\label{fVectors}
\end{figure}

Please see \cite{srivastava2013mixed} for a comprehensive description of the formulation. In this paper we present a summary of the same for completeness. In the following treatment repeated Latin indices mean summation, whereas, repeated Greek indices do not. Consider the problem of elastic wave propagation in a general 3-dimensional periodic composite. The unit cell of the periodic composite is denoted by $\Omega$ and is characterized by 3 base vectors $\mathbf{h}^i$, $i=1,2,3$. Any point within the unit cell can be uniquely specified by the vector $\mathbf{x}=H_i\mathbf{h}^i$ where $0\leq H_i\leq 1,\forall i$. The same point can also be specified in the orthogonal basis as $\mathbf{x}=x_i\mathbf{e}^i$. The reciprocal base vectors of the unit cell are given by:
\begin{equation}
\mathbf{q}^1=2\pi\frac{\mathbf{h}^2\times\mathbf{h}^3}{\mathbf{h}^1\cdot(\mathbf{h}^2\times\mathbf{h}^3)};\quad \mathbf{q}^2=2\pi\frac{\mathbf{h}^3\times\mathbf{h}^1}{\mathbf{h}^2\cdot(\mathbf{h}^3\times\mathbf{h}^1)};\quad \mathbf{q}^3=2\pi\frac{\mathbf{h}^1\times\mathbf{h}^2}{\mathbf{h}^3\cdot(\mathbf{h}^1\times\mathbf{h}^2)}
\end{equation}
such that $\mathbf{q}^i\cdot\mathbf{h}^j=2\pi\delta_{ij}$. Fig. (\ref{fVectors}) shows the schematic of a 2-D unit cell, clearly indicating the unit cell basis vectors, the reciprocal basis vectors and the orthogonal basis vectors.

The wave vector for a Bloch-wave traveling in the composite are given as $\mathbf{k}=Q_i\mathbf{q}^i$ where $0\leq Q_i\leq 1,\forall i$. The composite is characterized by a spatially varying stiffness tensor, $C_{jkmn}(\mathbf{x})$, and density, $\rho(\mathbf{x})$, which satisfy the following periodicity conditions:
\begin{equation}
C_{jkmn}(\mathbf{x}+n_i\mathbf{h}^i)=C_{jkmn}(\mathbf{x});\quad \rho(\mathbf{x}+n_i\mathbf{h}^i)=\rho(\mathbf{x})
\end{equation}
where $n_i$$(i=1,2,3)$ are integers.

\subsection{Field equations and boundary conditions}

For harmonic elastodynamic problems the equations of motion and kinematic relations at any point $\mathbf{x}$ in $\Omega$ are given by
\begin{equation}\label{equationofmotion}
\sigma_{jk,k}=-\lambda\rho u_j; \quad \varepsilon_{jk}=.5(u_{j,k}+u_{k,j})
\end{equation}
where $\lambda=\omega^2$, and $\boldsymbol{\sigma}e^{-i\omega t},\boldsymbol{\varepsilon}e^{-i\omega t},\mathbf{u}e^{-i\omega t}$ are the space and time dependent stress tensor, strain tensor, and displacement vector respectively. The stress tensor is related to the strain tensor through the elasticity tensor, $\sigma_{jk}=C_{jkmn}\varepsilon_{mn}$. The traction and displacement at any point in the composite are related to the corresponding traction and displacement at another point, sperated from the first by a unit cell, through Bloch relations. These relations serve as the homogeneous boundary conditions on $\partial\Omega$. If the Bloch wave vector is $\mathbf{k}$ then these boundary conditions are given by:
\begin{equation}\label{boundaryconditions}
u_j(\mathbf{x}+\mathbf{h}^i)=u_j(\mathbf{x})e^{i\mathbf{k}\cdot\mathbf{h}^i};\quad t_j(\mathbf{x}+\mathbf{h}^i)=-t_j(\mathbf{x})e^{i\mathbf{k}\cdot\mathbf{h}^i}, \quad \mathbf{x}\in\partial\Omega
\end{equation}
where $t_j=\sigma_{jk}\nu_k$ are the components of the traction vector and $\boldsymbol{\nu}$ is the exterior normal vector on $\partial\Omega$.

\subsection{Mixed-variational formulation}

It has been shown \cite{minagawa1976harmonic} that the solution to (\ref{equationofmotion}) that satisfies the boundary conditions, (\ref{boundaryconditions}), renders the following functional stationary:
\begin{equation}\label{mixedvariational}
\lambda_N=\frac{\langle\sigma_{jk},u_{j,k}\rangle+\langle u_{j,k},\sigma_{jk}\rangle+\langle D_{jkmn}\sigma_{jk},\sigma_{mn}\rangle}{\langle\rho u_j,u_j\rangle}
\end{equation}
where $\mathbf{D}$ is the compliance tensor and the inner product is given by:
\begin{equation}
\langle u,v\rangle=\int_\Omega uv^*d\Omega
\end{equation}
where $v^*$ is the complex conjugate of $v$.

\subsection{Approximation with periodic test functions}

We approximate the stress and displacement fields with the following test functions:
\begin{equation}\label{approximation}
\bar{u}_j=\sum_{\alpha,\beta,\gamma}U^{\alpha\beta\gamma}_jf^{\alpha\beta\gamma}(\mathbf{x}),\quad \bar{\sigma}_{jk}=\sum_{\alpha,\beta,\gamma}S^{\alpha\beta\gamma}_{jk}f^{\alpha\beta\gamma}(\mathbf{x})
\end{equation}
where the test functions satisfy the boundary conditions, (\ref{boundaryconditions}), and are orthogonal in the sense that $\langle f^{\alpha\beta\gamma},f^{\theta\eta\xi}\rangle$ is proportional to $\delta_{\alpha\theta}\delta_{\beta\eta}\delta_{\gamma\xi}$, $\boldsymbol{\delta}$ being the Kronecker delta. Substituting from (\ref{approximation}) to (\ref{mixedvariational}) and setting the derivative of $\lambda_N$ with respect to the unknown coefficients, ($U^{\alpha\beta\gamma}_j,S^{\alpha\beta\gamma}_{jk}$), equal to zero, we arrive at the following system of linear homogeneous equations:
\begin{align}\label{equationshomogeneous}
\langle\bar{\sigma}_{jk,k}+\lambda_N\rho\bar{u}_j,f^{\theta\eta\xi}\rangle=0\nonumber\\
\langle D_{jkmn}\bar{\sigma}_{mn}-\bar{u}_{(j,k)},f^{\theta\eta\xi}\rangle=0\nonumber\\
j,k,m,n=1,2,3
\end{align}
where $\bar{u}_{(j,k)}\equiv\bar{\varepsilon}_{jk}=.5(\bar{u}_{j,k}+\bar{u}_{k,j})$. For the general 3-dimensional case, if $\alpha,\beta,\gamma,\theta,\eta,\xi$ vary from $-M$ to $M$ then (\ref{equationshomogeneous}) represents $9(2M+1)^3$ linear homogeneous equations in the $9(2M+1)^3$ unknown displacement and stress coefficients. Given the symmetry of the stress tensor, these coefficients are $3(2M+1)^3$ number of $U^{\alpha\beta\gamma}_j$ and $6(2M+1)^3$ number of independent $S^{\alpha\beta\gamma}_{jk}$.

To approximate the stress and displacement fields in (\ref{approximation}), we use Fourier test functions of the following form:
\begin{equation}
f^{\alpha\beta\gamma}(\mathbf{x})=e^{i(\mathbf{k}\cdot\mathbf{x}+2\pi[\alpha H_1+\beta H_2+\gamma H_3])}
\end{equation}
where $\mathbf{x}=H_j\mathbf{h}^j$.

\section{Numerical solution}

The band-structure of the composite is given by the $\mathbf{q}-\omega$ pairs which lead to nontrivial solutions of (\ref{equationshomogeneous}). To calculate these pairs (\ref{equationshomogeneous}) is first written in the following equivalent matrix form:
\begin{align}\label{equationshomogeneousMatrix}
\nonumber \mathbf{HS}+\lambda_N\mathbf{\Omega U}=0\\
\mathbf{\Phi S}+\mathbf{H^*U}=0
\end{align}
Column vectors $\mathbf{S},\mathbf{U}$ contain the unknown coefficients of the periodic expansions of stress and displacement respectively. Matrices $\mathbf{H},\mathbf{\Omega},\mathbf{\Phi},\mathbf{H}^*$ contain the integrals of the various functions appearing in (\ref{equationshomogeneous}). Their sizes depend upon whether the problem under consideration is 1-, 2-, or 3-dimensional. These matrices would be described more clearly in the subsequent sections in which numerical examples are shown. The above system of equations can be recast into the following traditional eigenvalue problem:
\begin{equation}\label{eigenvalueproblem}
(\mathbf{H}\mathbf{\Phi}^{-1}\mathbf{H}^*)^{-1}\mathbf{\Omega}\mathbf{U}=\frac{1}{\lambda_N}\mathbf{U}
\end{equation}
whose eigenvalue solutions represent the frequencies ($\omega_N=\sqrt{\lambda_N}$) associated with the wave-vector under consideration ($\mathbf{q}$). The eigenvectors of the above equation are used to calculate the displacement modeshapes from (\ref{approximation}). The relation $\mathbf{S}=-\mathbf{\Phi}^{-1}\mathbf{H}^*\mathbf{U}$ is used to evaluate the stress eigenvector which is subsequently used to calculate the stress modeshape from (\ref{approximation}). The integrals occurring in (\ref{equationshomogeneousMatrix}) are numerically calculated over $\Omega$. Numerical integration is achieved by dividing the domain $\Omega$ into $P$ subdomains $\Omega_i,i=1,2...P$. The volume integral of any function $F(\mathbf{x})$ is then approximated as:
\begin{equation}
\int_\Omega F(\mathbf{x})d\Omega=\sum_i^PF_iV_i
\end{equation}
where $F_i$ is the value of the function $F(\mathbf{x})$ evaluated at the centroid of $\Omega_i$ and $V_i$ is the volume of $\Omega_i$. For meshing in 2-, and 3-D we have used a freely available Finite Element software \cite{geuzaine2009gmsh}.

\section{Serial calculation for 2-D periodic composites}

There are five possible Bravais lattices in 2 dimensions. However, they can be specified using two unit cell vectors ($\mathbf{h}^1,\mathbf{h}^2$). The reciprocal vector are $\mathbf{q}^1,\mathbf{q}^2$. The wave-vector of a Bloch wave traveling in this composite is specified as $\mathbf{k}=Q_1\mathbf{q}^1+Q_2\mathbf{q}^2$. To characterize the band-structure of the unit cell we evaluate the dispersion relation along the boundaries of the irreducible Brillouin zone. For purposes of demonstration and comparison we consider the case of plane strain state in the composite. The relevant stress components for the plane strain case are $\sigma_{11},\sigma_{22},\sigma_{12}$ and the relevant displacement components are $u_1,u_2$. The equations of motion and the constitutive law are:
\begin{equation}\label{equationofmotion2D}
\sigma_{jk,k}=-\lambda\rho(\mathbf{x}) u_j; \quad D_{jkmn}(\mathbf{x})\sigma_{mn}=u_{j,k};\quad j,k,m,n=1,2
\end{equation}
where $\mathbf{D}$ is the compliance tensor. For an isotropic material in plane strain $\mathbf{D}$ is given by:
\begin{equation}
D_{jkmn}=\frac{1}{2\mu}\left[\frac{1}{2}(\delta_{jm}\delta_{kn}+\delta_{jn}\delta_{km})-\frac{\lambda}{2(\mu+\lambda)}\delta_{jk}\delta_{mn}\right];\quad j,k,m,n=1,2
\end{equation}
where $\lambda,\mu$ are the Lame$^{'}$ constants of the material. The stresses and displacements are approximated by the following 2-D periodic functions:
\begin{equation}\label{approximation2d}
\bar{u}_j=\sum_{\alpha,\beta=-M}^MU^{\alpha\beta}_{j}e^{i2\pi Q^{\alpha\beta}_l x_l},\quad \bar{\sigma}_{jk}=\sum_{\alpha,\beta=-M}^MS^{\alpha\beta}_{jk}e^{i2\pi Q^{\alpha\beta}_l x_l};\quad j,k,l=1,2
\end{equation}
where
\begin{align}
\nonumber Q^{\alpha\beta}_1=T_{11}(Q_1+\alpha)+T_{21}(Q_2+\beta)\\
Q^{\alpha\beta}_2=T_{12}(Q_1+\alpha)+T_{22}(Q_2+\beta)
\end{align}
and the square matrix $[\mathbf{T}]$ is the inverse of the matrix $[\mathbf{A}]$ with components $[\mathbf{A}]_{jk}=\mathbf{h}^j\cdot\mathbf{e}^k$.

\subsection{Details of the matrices}

The matrix form of the eigenvalue problem is given by (\ref{equationshomogeneousMatrix}) with the following column vectors:
\begin{align}
\mathbf{U}=\{U^{\alpha\beta}_1\;U^{\alpha\beta}_2\}^T\nonumber\\
\mathbf{S}=\{S^{\alpha\beta}_{11}\;S^{\alpha\beta}_{22}\;S^{\alpha\beta}_{12}\}^T
\end{align}
Since $\alpha,\beta,\theta,\eta$ vary from $-M$ to $M$, the length of the column vector $\mathbf{U}$ is $2(2M+1)^2$ and the length of $\mathbf{S}$ is $3(2M+1)^2$. Corresponding to these column vectors, the size of $\mathbf{H}$ is $3(2M+1)^2\times 2(2M+1)^2$, $\mathbf{\Omega}$ is $2(2M+1)^2\times 2(2M+1)^2$, and $\mathbf{\Phi}$ is $3(2M+1)^2\times 3(2M+1)^2$. To clarify the contents of the matrices $[\mathbf{H}],[\mathbf{\Omega}],[\mathbf{\Phi}]$ we introduce the following modified coordinates:
\begin{align*}
I_1=(\alpha+M)(2M+1)+(\beta+1+M);\quad J_1=(\theta+M)(2M+1)+(\eta+1+M)\\
I_2=I_1+(2M+1)^2;\;J_2=J_1+(2M+1)^2\\
I_3=I_2+(2M+1)^2;\;J_3=J_2+(2M+1)^2
\end{align*}
Components of the $\mathbf{H}$ matrix are given by:
\begin{equation}\label{eh}
[\mathbf{H}]_{I_1J_1}=i2\pi Q^{\alpha\beta}_1\int_\Omega fd\Omega;\quad [\mathbf{H}]_{I_2J_2}=i2\pi Q^{\alpha\beta}_2\int_\Omega fd\Omega;\quad
[\mathbf{H}]_{I_1J_3}=[\mathbf{H}]_{I_2J_2};\quad [\mathbf{H}]_{I_2J_3}=[\mathbf{H}]_{I_1J_1}
\end{equation}
We also have $[\mathbf{H}]^*=-[\mathbf{H}]^T$ where the superscript $T$ denotes a matrix transpose. Components of the $\mathbf{\Omega}$ matrix are given by:
\begin{equation}\label{eo}
[\mathbf{\Omega}]_{I_1J_1}=\int_\Omega
\rho(x_1,x_2)fd\Omega;\quad[\mathbf{\Omega}]_{I_2J_2}=[\mathbf{\Omega}]_{I_1J_1}
\end{equation}
The rest of the terms in the $\mathbf{\Omega}$ matrix being zero. The components of the $\mathbf{\Phi}$ matrix are given by:
\begin{align}\label{ep}
\nonumber[\mathbf{\Phi}]_{I_1J_1}=\int_\Omega D_{1111}(x_1,x_2)fd\Omega\;\quad [\mathbf{\Phi}]_{I_1J_2}=\int_\Omega D_{1122}(x_1,x_2)fd\Omega;\quad [\mathbf{\Phi}]_{I_1J_3}=2\int_\Omega D_{1112}(x_1,x_2)fd\Omega\\
\nonumber[\mathbf{\Phi}]_{I_2J_1}=\int_\Omega D_{2211}(x_1,x_2)fd\Omega;\quad [\mathbf{\Phi}]_{I_2J_2}=\int_\Omega D_{2222}(x_1,x_2)fd\Omega;\quad [\mathbf{\Phi}]_{I_2J_3}=2\int_\Omega D_{2212}(x_1,x_2)fd\Omega\\
[\mathbf{\Phi}]_{I_3J_1}=2\int_\Omega D_{1211}(x_1,x_2)fd\Omega;\quad [\mathbf{\Phi}]_{I_3J_2}=2\int_\Omega D_{1222}(x_1,x_2)fd\Omega;\quad [\mathbf{\Phi}]_{I_3J_3}=4\int_\Omega D_{1212}(x_1,x_2)fd\Omega
\end{align}
where $f\equiv e^{i2\pi[(Q^{\alpha\beta}_1-Q^{\theta\eta}_1)x_1+(Q^{\alpha\beta}_2-Q^{\theta\eta}_2)x_2]}$. The above relations are provided for the most general case. For isotropic plane strain case several compliance components go to zero. This would result in several components of the $\mathbf{\Phi}$ matrix going to zero. Additionally, given the periodicity of the exponential function, several terms in $[\mathbf{H}],[\mathbf{H}]^*$ will also go to zero.

\subsection{Example: hexagonal unit cell}

\begin{figure}[htp]
\centering
\includegraphics[scale=.3]{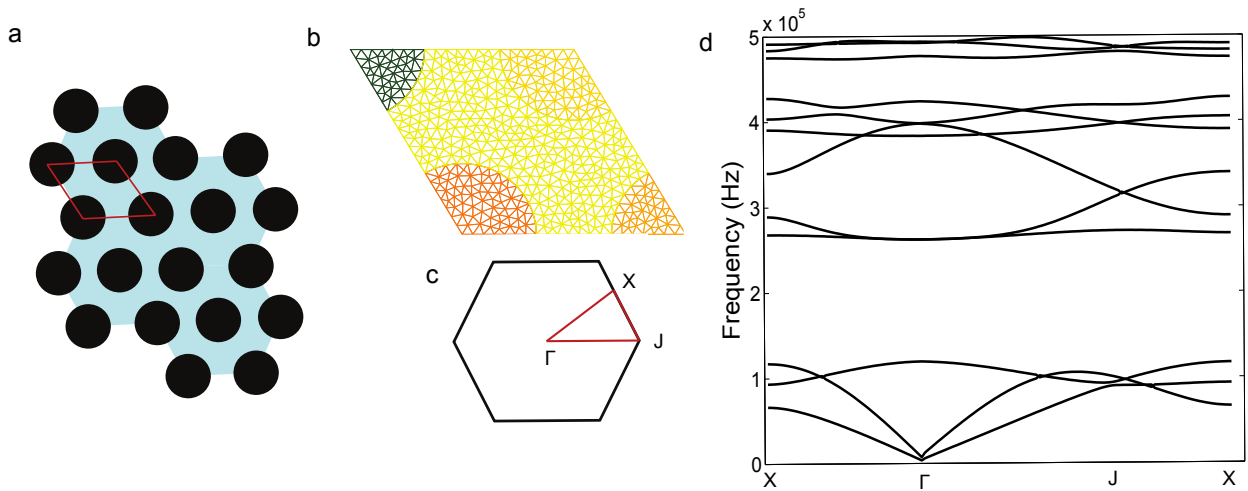}
\caption{a. Schematic of the 2-D periodic composite made from steel cylinders distributed in hexagonal packing in epoxy matrix, b. Discretization of the unit cell, c. Irreducible Brouillon Zone in the reciprocal lattice, d. Band-structure calculation results using the mixed variational formulation.}\label{fCompHexagonal}
\end{figure}

We consider a hexagonal unit cell made up of steel cylinders ordered in an epoxy matrix (Fig. \ref{fCompHexagonal}a). The diameter of the steel cylinders is 4mm and the lattice constant is 6.023mm. The material properties are taken from \cite{vasseur2001experimental} and are reproduced here for reference

\begin{enumerate}
\item Steel: $C_{11}=264$ Gpa, $C_{44}=81$ Gpa, $\rho=7780$ kg/m$^3$
\item Epoxy: $C_{11}=7.54$ Gpa, $C_{44}=1.48$ Gpa, $\rho=1142$ kg/m$^3$
\end{enumerate}

Fig. (\ref{fCompHexagonal}b) shows the automatic discretization of the unit cell into 964 triangular elements. The band-structure is evaluated along the boundaries of the Irreducible Brouillon zone. This boundary is denoted by the path $X-\Gamma-J-X$ and is shown in Fig. (\ref{fCompHexagonal}c) in the reciprocal cell. For the band-structure calculations we use a total of 242 terms ($M=5$). This results in the simultaneous evaluation of the first 242 eigenvalues for each wavenumber point. The results in Fig. (\ref{fCompHexagonal}d), however, only show the first 12 eigenvalues. These results are in very good agreement with the PWE calculations shown in Ref.\cite{vasseur2001experimental} (Fig. 3 in that paper). We note the existence of the all-angle stop-band for waves traveling in the plane of the unit cell in the frequency ranges of 120-262 kHz and 427-473 kHz. The locations of the stop-bands and the general shape of the pass-bands are shown to match very well with the results in Ref.\cite{vasseur2001experimental}.

\subsection{Comments on the efficiency of serial computations}

We note that the central computation which is being performed in the formulation presented above is the calculation of the $f$ integrals over the geometry of the unit cell. These integrals are shown more clearly in equations (\ref{eh},\ref{eo},\ref{ep}) for the 2-D case. The integrands in each of these equations depend upon the Fourier terms $\alpha,\beta,\theta,\eta$ and/or the coordinates of the points within the unit cell. There appears no easy way of computing these integrals once and using the saved value for further calculations. The most straightforward way of implementing integrals is to evaluate the integrands for each set of $\alpha,\beta,\theta,\eta$ as they assume values from $-M,...M$ and for each element-centroid point in the mesh. These computations present a significant bottleneck in the formulation. For M=6 and with thousand elements in the mesh, each integral requires the evaluation of its integrand at $1000\times (2M+1)^4$ different points or at 28.561 million points. To calculate all the integrals in the equations (assuming that the symmetry of the compliance matrix renders some calculations redundant) we require the evaluation of the integrands at 228.488 million points. In addition to this significantly time consuming step, the formulation requires further matrix manipulations to calculate $(2M+1)^2$ eigen-frequency values for a given wave-vector point. To calculate the entire band-structure along the boundaries of the IBZ, it is required to discretize the boundary at several different wave-vector points and to use the formulation at each of the discretized points. If the boundary is dicretized at 70 wave-vector points, the mixed-variational formulation requires the calculation of the integrands 16 billion times.
\begin{figure}[htp]
\centering
\includegraphics[scale=.5]{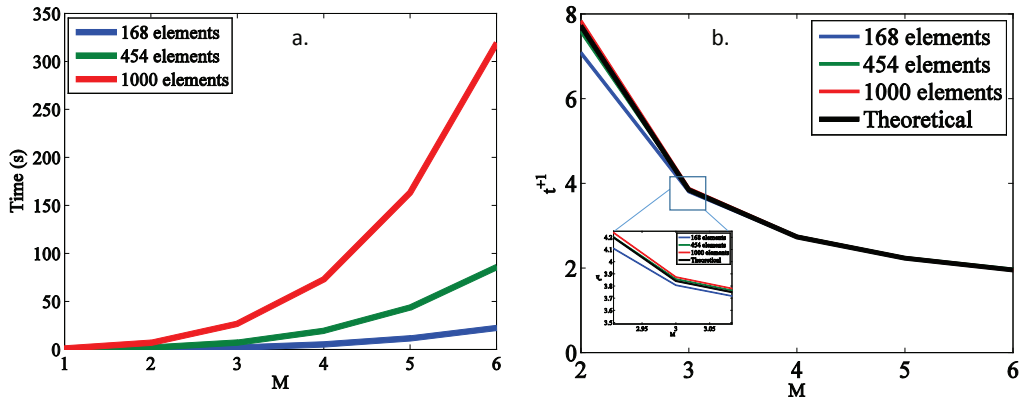}
\caption{a. Time in seconds taken by mixed-variation MATLAB code to calculate the eigen-frequencies at one wave-vector point ($Q_1=.25,Q_2=0$), b. $t^{+1}$ for $m'$ from 2 to 6.}\label{fMATLABtimes}
\end{figure}
To illustrate the considerable time required to achieve this through traditional loops we present eigen-frequency computation times (averaged over 5 runs) for one wave-vector point through a MATLAB implemented code (Fig. \ref{fMATLABtimes}a). The implementation is serial with loops being employed to cycle over the Fourier terms and the elements in the mesh. The results are shown for 3 different levels of mesh discretization and 5 different values of $M$. For $M=1$ and coarse unit cell discretization (168 elements) MATLAB returns the results in an average of .07 seconds. However, such a coarse computation gives unreliable results for the higher branches. To reliably evaluate higher branches we require more Fourier terms and a finer mesh dicretization to accurately represent the spatial variations of the field variables over the unit cell. This increases the computation times significantly. An approximate estimate may be obtained for the time required in serial computations for finer meshes and higher $M$ values. If the time taken by the serial computations for $n$ elements in the mesh and for $M=m$ is $t_{nm}$ then we should have
\begin{equation}
t_{n',n}\equiv\frac{t_{n'm}}{t_{nm}}\approx\frac{n}{n'};\quad t^{m',m}\equiv\frac{t_{nm'}}{t_{nm}}\approx\frac{(2m'+1)^4}{(2m+1)^4}
\end{equation}
The approximation results from considering only the assembly of the matrices as the dominant computation. Therefore the approximation should improve for high values of $n,n',m,m'$ where loop processes dominate. There is a linear increase in the computation time with increase in the number of elements but a power-4 increase in time with $M$ values. Denoting $t^{m',m'-1}$ by $t^{+1}$ we note that $t^{+1}$ is always greater than 1 but converges to 1 for large $m'$. This parameter is plotted in Fig. \ref{fMATLABtimes}b for $m'$ from 2 to 6. It can be seen that the actual computation time factors are very close to the theoretical factors, attesting to the initial assumption that it is the assembly of the matrices which is the dominant computation in the method. It is clear that the power-4 time complexity involved with $M$ in the 2-D case (and the power-6 time complexity in the 3-D case) renders the application of the method slow for higher branches. For instance, 318.9638 seconds are required to compute the eigen-frequencies for $M=6$ and with a mesh discretization of 1000 elements. Since the results shown in Fig. \ref{fMATLABtimes}a pertain to 1 wave-vector point, a rough estimate of the time required to solve the complete band-structure of a 2-D composite may be made by multiplying with the number of distinct wave-vector points at which the computation is desired. If the boundary of the IBZ is discretized at 70 different wave-vector points, it would tentatively require 6.2 hours to do the complete calculation for $M=6$ and for 1000 elements in the mesh. Although some efficiency  gains may be achieved on regular cpus by exploiting their multi-core, multi-threaded nature, the time required for the computation would still be far too prohibitive to use the formulation in areas like band-structure optimization and inverse problems in dynamic homogenization where thousands of unit cells potentially need to be analyzed.

In the present form of the formulation the maximum parallelization that can be achieved is at the level of wave-vector discretization. Given 70 parallel cpu cores each of which is comparable to the the present hardware, it should be possible to analyze the complete band-structure of the 2-D unit cell (at 70 points on the boundary of the IBZ) in around 318 seconds. However 70 parallel cpu cores represent a significant financial investment. Moreover, incorporating further parallelization appears difficult even with the availability of more parallel processing capabilities. Our aim in this paper is to present the formulation in a form which is not limited in its parallelizability. Moreover we present it within the context of gpu-acceleration. Since gpus are much cheaper per parallel thread compared to cpus, the formulation  presented in this paper can be used for very cost-effective computations.

\section{gpu-computations}

In the present problem we note that even though billions of computations are required for the complete solution of the band-structure, all of these computations are themselves very simple. More importantly, all of these computations are independent of each other and hence can be carried out in parallel if the formulation can be properly recast. The ideal requirement from the computational platform, therefore, is that of massive parallel capability. The highest end desktop cpu processors currently available have 8 independent cores capable of running 16 independent processes in parallel through hyperthreading. Further parallelization may be achieved by assembling clusters of such processors together effectively resulting in many parallel computational cores but it requires significant financial investment. Graphical processors, which are normally used to render graphics on modern laptops and desktops, provide a cheap and efficient alternative to cpu processing when a large number of parallel processes are required to be computed.

We have incorporated the mixed-variational formulation in python and used the CUDA parallel computing platform to execute the most computationally intensive parts of the formulation on an NVIDIA Quadro K2000 gpu. The K2000 has a total of 384 CUDA cores each of which can execute multiple processes in parallel. gpus with more cores are available for cheap and the formulation is presented in a way which makes it easily applicable to higher levels of parallelization. To achieve such parallelization we have recast the algorithm so as to replace almost all loop operations with equivalent matrix operations. One of the basic computational unit in the formulation is the calculation of the following $f$ integral over the unit cell (for $\mathbf{H}$ matrix in \ref{eh}):
\begin{equation}
I^{(1)}_{\alpha\beta\theta\eta}=\int_\Omega fd\Omega=\int_\Omega e^{i2\pi[(Q^{\alpha\beta}_1-Q^{\theta\eta}_1)x_1+(Q^{\alpha\beta}_2-Q^{\theta\eta}_2)x_2]}d\Omega\equiv\sum_{i=1}^{N_{el}}f_i(\alpha,\beta,\theta,\eta)A_i
\end{equation}
where $f_i$ is the evaluation of the integrand at the centroid location of the $i^{th}$ element in the mesh and $A_i$ is the area of the $i^{th}$ element. The calculation of the integral for different values of $\alpha,\beta,\theta,\eta$, if implemented in the most straightforward way possible, represents $N_{el}\times(2M+1)^4$ loops within which the integrand needs to be evaluated. To expedite the computations we view the result of all the evaluations of the integrand at all Fourier points as a 3-dimensional matrix $[f]$ with size [$N_{el},(2M+1)^2,(2M+1)^2$]. Each element [$i,j,k$] of this matrix is evaluated from one 2-dimensional matrix and four 1-dimensional vectors. The first matrix of size [$N_{el},2$], denoted by $[c]$, represents the centroid values of the mesh. Its [$i,1$] and [$i,2$] elements represent the $x_1$ and $x_2$ coordinates, respectively, of the centroid of the $i^{th}$ mesh-element. The second and third vectors ($[A_1],[A_2]$) of length $(2M+1)^2$ represents the appropriate combination of $\alpha,\beta$ occurring in the integral. They are derived by transforming two 2-dimensional matrices of size [$2M+1,2M+1$] whose [$m,n$] elements are respectively given by:
\begin{eqnarray}
\nonumber i2\pi\left[T(0,0)(Q_1+\alpha(m)) + T(0,1)(Q_2+\beta(n))\right]\quad\quad\text{for $A_1$}\\
i2\pi\left[T(1,0)(Q_1+\alpha(m)) + T(1,1)(Q_2+\beta(n))\right]\quad\quad\text{for $A_2$}
\end{eqnarray}
where $\alpha,\beta=-M...M$ are two 1-dimensional vectors of length $2M+1$. Similarly the fourth and fifth vectors ($[B_1],[B_2]$) of length $(2M+1)^2$ represents the appropriate combination of $\theta,\eta$ appearing in the integral. They are derived by transforming two 2-dimensional matrices of size [$2M+1,2M+1$] whose [$m,n$] elements are respectively given by:
\begin{eqnarray}
\nonumber 2i\pi\left[T(0,0)(Q_1+\theta(m)) + T(0,1)(Q_2+\eta(n))\right]\quad\quad\text{for $B_1$}\\
2i\pi\left[T(1,0)(Q_1+\theta(m)) + T(1,1)(Q_2+\eta(n))\right]\quad\quad\text{for $B_2$}
\end{eqnarray}
where $\theta,\eta=-M...M$ are two 1-dimensional vectors of length $2M+1$. Expressed in this way the [$i,j,k$] value of the integrand matrix (denoted by $[f]_{ijk}$) is given by:
\begin{equation}\label{eExp}
[f]_{ijk}=e^{\left[A_1(j)-B_1(k)\right]c(i,1)+\left[A_2(j)-B_2(k)\right]c(i,2)}
\end{equation}
with the integral given by:
\begin{equation}
I^{(1)}_{jk}=\sum_{i=1}^{N_{el}}[f]_{ijk}A_i;\quad j,k=1,2,...(2M+1)^2
\end{equation}
To compute this matrix of the integrals the matrices $c,A_1,A_2,B_1,B_2,A$ are passed to the gpu. On the gpu the computational kernels are executed by a grid of thread blocks, where each block is a grid of threads. Each thread has a unique id which can be used to map it to a unique set of indices $i,j,k$. Each thread then performs the simple calculations in (\ref{eExp}) using the vector and matrix elements which are uniquely identified by the set $i,j,k$. Since the actual computation carried out by each thread is relatively simple and many threads are running in parallel, the method displays considerably reduced calculation times in comparison with the straightforward loops method. The integral appearing in \ref{eo} is similarly given by
\begin{equation}
I^{(2)}_{jk}\equiv\int_\Omega
\rho(x_1,x_2)fd\Omega\equiv\sum_{i=1}^{N_{el}}\rho_i[f]_{ijk}A_i;\quad j,k=1,2,...(2M+1)^2
\end{equation}
where the vector $[\rho]$ contains the density information of the elements. The integrals in \ref{ep} can be similarly calculated. Through the highly parallel implementation of the mixed-variational formulation we report more than hundred fold efficiency gains in the mixed-variational band-structure algorithm, especially for large $M$ values.

\subsection{gpu-computations: results}

\begin{figure}[htp]
\centering
\includegraphics[scale=.5]{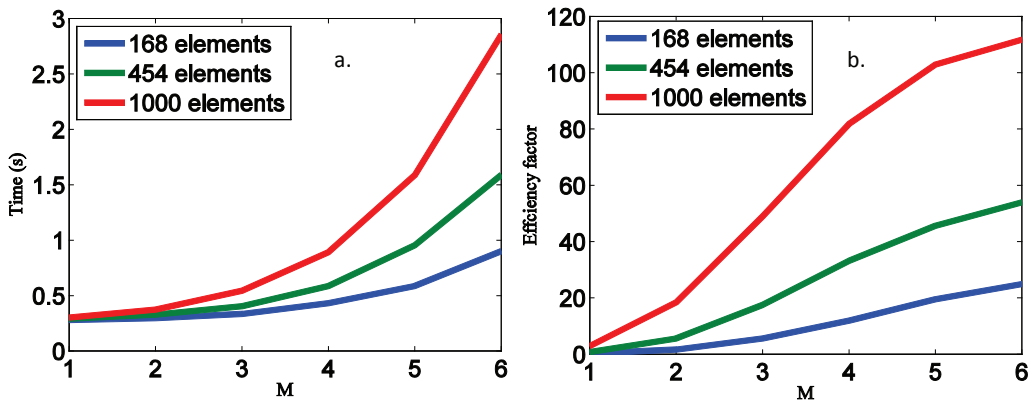}
\caption{a. Time in seconds taken by the gpu-accelerated mixed-variational formulation to calculate the eigen-frequencies at one wave-vector point ($Q_1=.25,Q_2=0$) b. Efficiency factor comparing the parallel formulation with the serial formulation}\label{fGPUtimes}
\end{figure}
To compare the gpu-computation results with the cpu results shown in Fig. \ref{fMATLABtimes}a, we present in Fig. \ref{fGPUtimes}a the times taken for the same computations performed on the same desktop computer. In the present case most of the computationally intensive tasks are performed on the gpu. Each point on Fig. \ref{fGPUtimes}a is an average of 5 runs. We define an efficiency factor which measures the performance improvement through the parallel computations over serial computations in terms of the time it takes to do the same computation through the two methods:
\begin{equation}
e=\frac{t_{serial}}{t_{parallel}}
\end{equation}
These efficiency factors are plotted in Fig. \ref{fGPUtimes}b. For $M=1$ the gpu computations take longer than the same cpu computations. This can be attributed to the time overhead involved in passing the matrices from the host (cpu) to the device (gpu). For $M=1$ the efficiency gained in the parallel evaluation of the integrals is lower than the efficiency lost in passing the required matrices to the gpu. However, for all other values of $M$ and for all levels of discretizations considered the parallel computations are faster than their serial counterparts. At $M=6$, the gpu performance is 25 times faster than the cpu computations for the coarse mesh (168 elements). For the fine mesh with 1000 elements the gpu computations are almost 112 times faster than cpu computations.

\subsection{Comments on effciency}

\begin{figure}[htp]
\centering
\includegraphics[scale=.5]{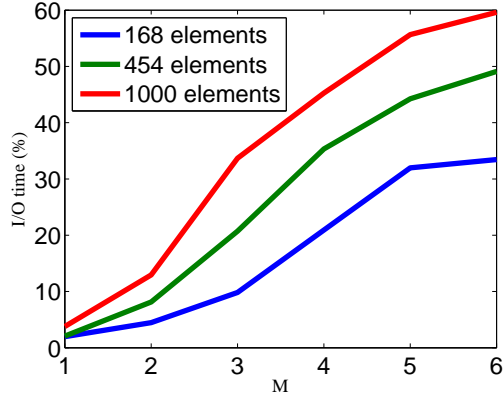}
\caption{Time taken in transferring data between the host and the device as a percentage of total execution time}\label{fMemoryTimes}
\end{figure}
Since part of the formulation presented above is executed on the cpu and part of it is executed on the gpu it is not easy to establish a theoretical estimate of the total execution time. However, in this section we show that the gpu-formulation is I/O bound as opposed to compute bound. While the serial version of the formulation was dominated by the assembly of the matrices which dictated the execution time, the parallel version is overwhelmingly dominated by the time it takes to pass data between the host and the device. The time taken in the assembly of the matrices is a negligibly small fraction of the total compute time. Fig. (\ref{fMemoryTimes}) shows the total time taken in transferring the data between the host and the device as a percentage of the total execution time for three mesh discretization values and for $M=1,2..,6$. For $M=1$ and the I/O time is less than $5\%$ of the total execution time for all three levels of discretization. However, the I/O time accounts for more than $50\%$ of the total execution time for $M=6$ and the fine mesh. Even though it takes billions (if not trillions) of floating point operations to assemble the matrices for $M=6$ and 1000 elements, the time taken by the gpu for the assembly is less than $.02\%$ of the total execution time. This translates into an assembly time of less than .57 milliseconds suggesting the potential for further reduction in time if the I/O operations are optimized. In view of the small compute times, this result becomes particularly significant for cases where thousands of micro-structures may need to be analyzed for the optimization of some dynamic property.

\subsection{Full band-structure calculations}
\begin{figure}[htp]
\centering
\includegraphics[scale=.3]{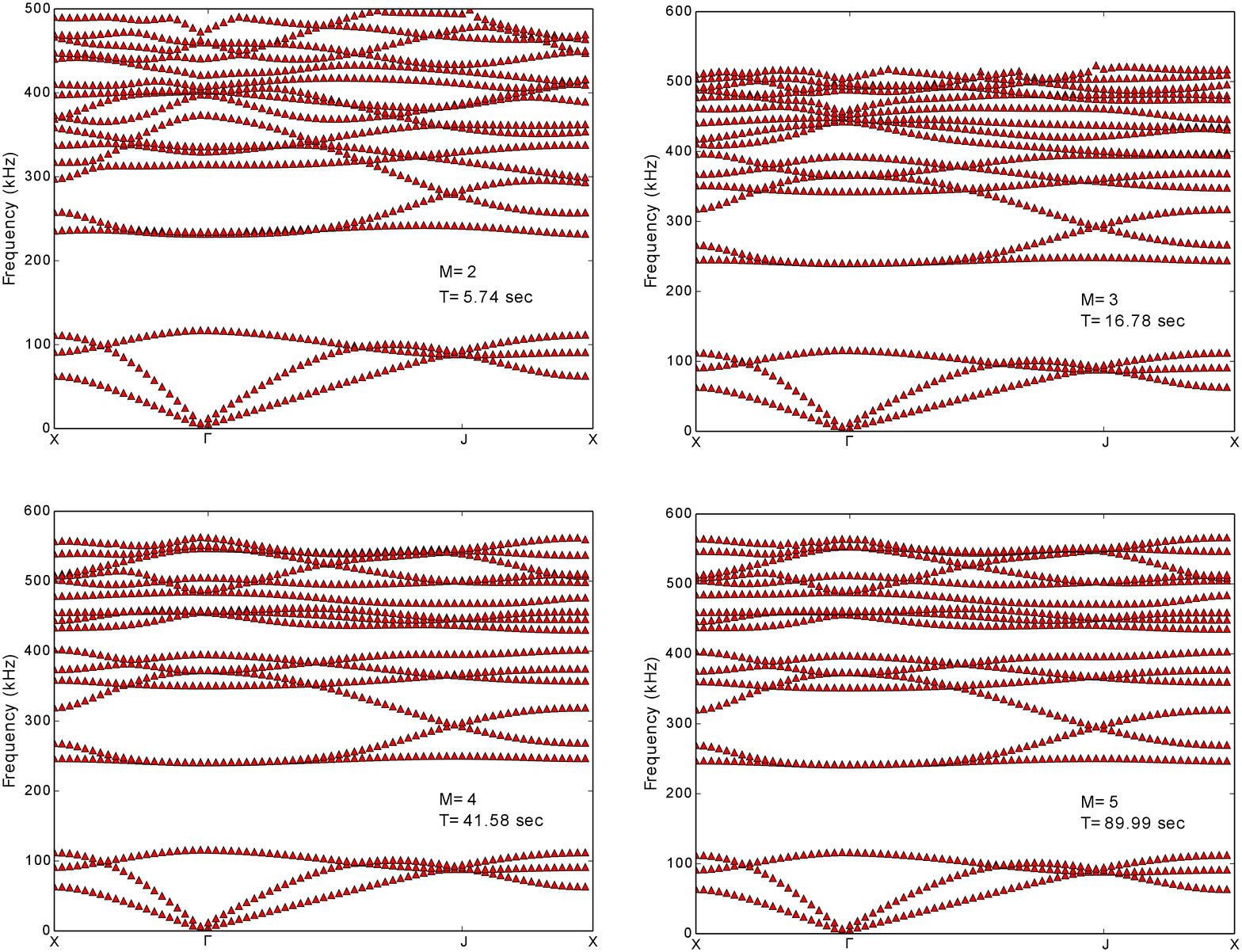}
\caption{Full band-structure calculation through gpu computing. The boundary of the IBZ is discretized at 70 points. A total of 1180 elements are used in the computations.}\label{fGPUFullBand}
\end{figure}

The speed of the computation becomes more apparent when the full band-structure is evaluated through gpu-computing. Fig. \ref{fGPUFullBand} shows the band-structure of the 2-D composite along the boundary of the IBZ and for the first 18 branches. For these calculations we have discretized the boundary of the IBZ at 70 wave-vector points and used a mesh with 1000 elements. A similar computation on a cpu with rudimentary loops is estimated to take 3 hours when $M=5$. The gpu accelerated formulation provides the complete set of results in 90 seconds.
\begin{figure}[htp]
\centering
\includegraphics[scale=.35]{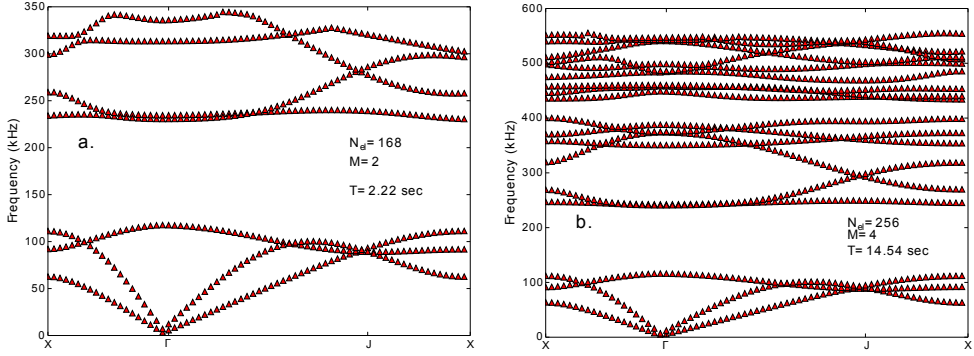}
\caption{Full band-structure calculation through gpu computing. The boundary of the IBZ is discretized at 70 points.}\label{fM2and4}
\end{figure}
From Fig. \ref{fGPUFullBand} it is also evident that for the first 18 branches the solution has acceptably converged even for $M=4$, a calculation which only takes 42 seconds. Furthermore, the solution computed for $M=4$ when 256 elements are used in the mesh gives accurate results for the first 18 branches and takes less than 15 seconds (Fig. \ref{fM2and4}b). Making the mesh more coarse or using less number of Fourier terms in the mixed-variational formulation results in worsening accuracy in the calculation of the branches. However, if only the first few branches are of interest then the gpu implementation of the mixed-variational formulation provides acceptable results in around 2 seconds (Fig. \ref{fM2and4}a).
\begin{figure}[htp]
\centering
\includegraphics[scale=.45]{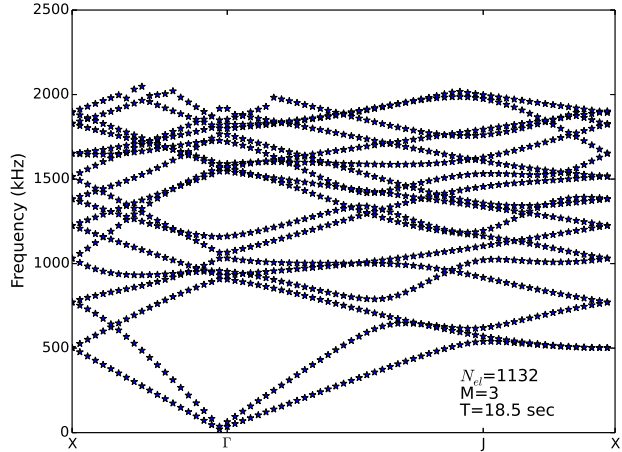}
\caption{Full band-structure calculation of a hexagonal unit cell with 1132 different phases.}\label{fArbitrary}
\end{figure}

To show that the gpu-accelerated mixed-variational method presented in this paper provides a fast band-structure computational tool for unit cells of arbitrary complexity we present the computed results for a hypothetical hexagonal unit cell divided into 1132 mesh-elements where each element has different material properties. The geometry of the unit cell is given in Fig. \ref{fCompHexagonal}a except that instead of being made up of 2 phases, the unit cell is made up of possibly 1132 different phases. While the stiffness properties for each phase are taken to be equal ($C_{11}=264$ Gpa, $C_{44}=81$ Gpa), the density of each phase is computed from the $x_2$ coordinate of the centroid of its mesh-element ($\rho_i=1000000\times[c]_{i,2}$). The band-structure of this composite is calculated from the formulation presented in this paper and is shown in Fig. \ref{fArbitrary}. For this case the solution shows convergence at $M=3$, a calculation which is performed in 18.5 seconds.

\section{Discussions and Conclusions}

In this paper we have presented a novel implementation of the mixed-variational method within the framework of gpu computing. We report more than 100 fold performance gains through gpu computing over the unoptimized serial implementation on a cpu. We showed, by way of an example, that a single mid-range gpu (K2000 in the present case) is powerful enough to solve the band-structure of a 2-D phononic composite in a matter of seconds. For the example chosen in this paper, it was shown that gpu computations result in acceptable band-structure calculations for the first 18 branches in less than 15 seconds. If only the first few branches are desired then the gpu-accelerated mixed-variational formulation provides the result in around 2 seconds. Finally we presented the application of the gpu-accelerated mixed-variational formulation to a hexagonal unit cell comprising of 1132 different phases. The speed of the results and the versatility of the method show its strong potential in the areas of band-structure optimization.

We also note that the gpu computations are I/O bound indicating the possibility of further efficiency through the use of optimized data transfer between the host and the device. Additionally we note that the system on which the computations were carried out is not an expensive cluster but a regular desktop computer. The gpu used in the calculations (K2000) is not the highest end gpu currently available. Additionally memory optimizations which can further accelerate the algorithm were not implemented in this paper. By using parallel gpus with more computational cores, optimized memory management schemes, and precomputed results wherever possible, it should be possible to further decrease the calculation times considerably. The current gpu was also limited in terms of its internal memory (2 GB) which prevented very large matrices from being manipulated. Such matrices are required for solving the phononic band-structures of 3-D composites. However, with more gpu memory it is expected that a mixed-variational implementation on distributed gpus can efficiently characterize the phononic properties of arbitrarily complex 3-D unit cells.

%\bibliography{../References/ReferencesBib}

\begin{thebibliography}{36}%
\makeatletter
\providecommand \@ifxundefined [1]{%
 \@ifx{#1\undefined}
}%
\providecommand \@ifnum [1]{%
 \ifnum #1\expandafter \@firstoftwo
 \else \expandafter \@secondoftwo
 \fi
}%
\providecommand \@ifx [1]{%
 \ifx #1\expandafter \@firstoftwo
 \else \expandafter \@secondoftwo
 \fi
}%
\providecommand \natexlab [1]{#1}%
\providecommand \enquote  [1]{``#1''}%
\providecommand \bibnamefont  [1]{#1}%
\providecommand \bibfnamefont [1]{#1}%
\providecommand \citenamefont [1]{#1}%
\providecommand \href@noop [0]{\@secondoftwo}%
\providecommand \href [0]{\begingroup \@sanitize@url \@href}%
\providecommand \@href[1]{\@@startlink{#1}\@@href}%
\providecommand \@@href[1]{\endgroup#1\@@endlink}%
\providecommand \@sanitize@url [0]{\catcode `\\12\catcode `\$12\catcode
  `\&12\catcode `\#12\catcode `\^12\catcode `\_12\catcode `\%12\relax}%
\providecommand \@@startlink[1]{}%
\providecommand \@@endlink[0]{}%
\providecommand \url  [0]{\begingroup\@sanitize@url \@url }%
\providecommand \@url [1]{\endgroup\@href {#1}{\urlprefix }}%
\providecommand \urlprefix  [0]{URL }%
\providecommand \Eprint [0]{\href }%
\providecommand \doibase [0]{http://dx.doi.org/}%
\providecommand \selectlanguage [0]{\@gobble}%
\providecommand \bibinfo  [0]{\@secondoftwo}%
\providecommand \bibfield  [0]{\@secondoftwo}%
\providecommand \translation [1]{[#1]}%
\providecommand \BibitemOpen [0]{}%
\providecommand \bibitemStop [0]{}%
\providecommand \bibitemNoStop [0]{.\EOS\space}%
\providecommand \EOS [0]{\spacefactor3000\relax}%
\providecommand \BibitemShut  [1]{\csname bibitem#1\endcsname}%
\let\auto@bib@innerbib\@empty
%</preamble>
\bibitem [{\citenamefont {Cervera}\ \emph {et~al.}(2001)\citenamefont
  {Cervera}, \citenamefont {Sanchis}, \citenamefont {Sanchez-Perez},
  \citenamefont {Martinez-Sala}, \citenamefont {Rubio}, \citenamefont
  {Meseguer}, \citenamefont {Lopez}, \citenamefont {Caballero},\ and\
  \citenamefont {S{\'a}nchez-Dehesa}}]{cervera2001refractive}%
  \BibitemOpen
  \bibfield  {author} {\bibinfo {author} {\bibfnamefont {F.}~\bibnamefont
  {Cervera}}, \bibinfo {author} {\bibfnamefont {L.}~\bibnamefont {Sanchis}},
  \bibinfo {author} {\bibfnamefont {J.}~\bibnamefont {Sanchez-Perez}}, \bibinfo
  {author} {\bibfnamefont {R.}~\bibnamefont {Martinez-Sala}}, \bibinfo {author}
  {\bibfnamefont {C.}~\bibnamefont {Rubio}}, \bibinfo {author} {\bibfnamefont
  {F.}~\bibnamefont {Meseguer}}, \bibinfo {author} {\bibfnamefont
  {C.}~\bibnamefont {Lopez}}, \bibinfo {author} {\bibfnamefont
  {D.}~\bibnamefont {Caballero}}, \ and\ \bibinfo {author} {\bibfnamefont
  {J.}~\bibnamefont {S{\'a}nchez-Dehesa}},\ }\href@noop {} {\bibfield
  {journal} {\bibinfo  {journal} {Physical review letters}\ }\textbf {\bibinfo
  {volume} {88}},\ \bibinfo {pages} {23902} (\bibinfo {year}
  {2001})}\BibitemShut {NoStop}%
\bibitem [{\citenamefont {Yang}\ \emph {et~al.}(2002)\citenamefont {Yang},
  \citenamefont {Page}, \citenamefont {Liu}, \citenamefont {Cowan},
  \citenamefont {Chan},\ and\ \citenamefont {Sheng}}]{yang2002ultrasound}%
  \BibitemOpen
  \bibfield  {author} {\bibinfo {author} {\bibfnamefont {S.}~\bibnamefont
  {Yang}}, \bibinfo {author} {\bibfnamefont {J.}~\bibnamefont {Page}}, \bibinfo
  {author} {\bibfnamefont {Z.}~\bibnamefont {Liu}}, \bibinfo {author}
  {\bibfnamefont {M.}~\bibnamefont {Cowan}}, \bibinfo {author} {\bibfnamefont
  {C.}~\bibnamefont {Chan}}, \ and\ \bibinfo {author} {\bibfnamefont
  {P.}~\bibnamefont {Sheng}},\ }\href@noop {} {\bibfield  {journal} {\bibinfo
  {journal} {Physical review letters}\ }\textbf {\bibinfo {volume} {88}},\
  \bibinfo {pages} {104301} (\bibinfo {year} {2002})}\BibitemShut {NoStop}%
\bibitem [{\citenamefont {Khelif}\ \emph {et~al.}(2003)\citenamefont {Khelif},
  \citenamefont {Choujaa}, \citenamefont {Djafari-Rouhani}, \citenamefont
  {Wilm}, \citenamefont {Ballandras},\ and\ \citenamefont
  {Laude}}]{khelif2003trapping}%
  \BibitemOpen
  \bibfield  {author} {\bibinfo {author} {\bibfnamefont {A.}~\bibnamefont
  {Khelif}}, \bibinfo {author} {\bibfnamefont {A.}~\bibnamefont {Choujaa}},
  \bibinfo {author} {\bibfnamefont {B.}~\bibnamefont {Djafari-Rouhani}},
  \bibinfo {author} {\bibfnamefont {M.}~\bibnamefont {Wilm}}, \bibinfo {author}
  {\bibfnamefont {S.}~\bibnamefont {Ballandras}}, \ and\ \bibinfo {author}
  {\bibfnamefont {V.}~\bibnamefont {Laude}},\ }\href@noop {} {\bibfield
  {journal} {\bibinfo  {journal} {physical Review B}\ }\textbf {\bibinfo
  {volume} {68}},\ \bibinfo {pages} {214301} (\bibinfo {year}
  {2003})}\BibitemShut {NoStop}%
\bibitem [{\citenamefont {Reed}\ \emph {et~al.}(2003)\citenamefont {Reed},
  \citenamefont {Solja{\v{c}}i{\'c}},\ and\ \citenamefont
  {Joannopoulos}}]{reed2003reversed}%
  \BibitemOpen
  \bibfield  {author} {\bibinfo {author} {\bibfnamefont {E.}~\bibnamefont
  {Reed}}, \bibinfo {author} {\bibfnamefont {M.}~\bibnamefont
  {Solja{\v{c}}i{\'c}}}, \ and\ \bibinfo {author} {\bibfnamefont
  {J.}~\bibnamefont {Joannopoulos}},\ }\href@noop {} {\bibfield  {journal}
  {\bibinfo  {journal} {Physical review letters}\ }\textbf {\bibinfo {volume}
  {91}},\ \bibinfo {pages} {133901} (\bibinfo {year} {2003})}\BibitemShut
  {NoStop}%
\bibitem [{\citenamefont {Yang}\ \emph {et~al.}(2004)\citenamefont {Yang},
  \citenamefont {Page}, \citenamefont {Liu}, \citenamefont {Cowan},
  \citenamefont {Chan},\ and\ \citenamefont {Sheng}}]{yang2004focusing}%
  \BibitemOpen
  \bibfield  {author} {\bibinfo {author} {\bibfnamefont {S.}~\bibnamefont
  {Yang}}, \bibinfo {author} {\bibfnamefont {J.}~\bibnamefont {Page}}, \bibinfo
  {author} {\bibfnamefont {Z.}~\bibnamefont {Liu}}, \bibinfo {author}
  {\bibfnamefont {M.}~\bibnamefont {Cowan}}, \bibinfo {author} {\bibfnamefont
  {C.}~\bibnamefont {Chan}}, \ and\ \bibinfo {author} {\bibfnamefont
  {P.}~\bibnamefont {Sheng}},\ }\href@noop {} {\bibfield  {journal} {\bibinfo
  {journal} {Physical review letters}\ }\textbf {\bibinfo {volume} {93}},\
  \bibinfo {pages} {24301} (\bibinfo {year} {2004})}\BibitemShut {NoStop}%
\bibitem [{\citenamefont {Gorishnyy}\ \emph {et~al.}(2005)\citenamefont
  {Gorishnyy}, \citenamefont {Ullal}, \citenamefont {Maldovan}, \citenamefont
  {Fytas},\ and\ \citenamefont {Thomas}}]{gorishnyy2005hypersonic}%
  \BibitemOpen
  \bibfield  {author} {\bibinfo {author} {\bibfnamefont {T.}~\bibnamefont
  {Gorishnyy}}, \bibinfo {author} {\bibfnamefont {C.}~\bibnamefont {Ullal}},
  \bibinfo {author} {\bibfnamefont {M.}~\bibnamefont {Maldovan}}, \bibinfo
  {author} {\bibfnamefont {G.}~\bibnamefont {Fytas}}, \ and\ \bibinfo {author}
  {\bibfnamefont {E.}~\bibnamefont {Thomas}},\ }\href@noop {} {\bibfield
  {journal} {\bibinfo  {journal} {Physical review letters}\ }\textbf {\bibinfo
  {volume} {94}},\ \bibinfo {pages} {115501} (\bibinfo {year}
  {2005})}\BibitemShut {NoStop}%
\bibitem [{\citenamefont {Mohammadi}\ \emph {et~al.}(2008)\citenamefont
  {Mohammadi}, \citenamefont {Eftekhar}, \citenamefont {Khelif}, \citenamefont
  {Hunt},\ and\ \citenamefont {Adibi}}]{mohammadi2008evidence}%
  \BibitemOpen
  \bibfield  {author} {\bibinfo {author} {\bibfnamefont {S.}~\bibnamefont
  {Mohammadi}}, \bibinfo {author} {\bibfnamefont {A.}~\bibnamefont {Eftekhar}},
  \bibinfo {author} {\bibfnamefont {A.}~\bibnamefont {Khelif}}, \bibinfo
  {author} {\bibfnamefont {W.}~\bibnamefont {Hunt}}, \ and\ \bibinfo {author}
  {\bibfnamefont {A.}~\bibnamefont {Adibi}},\ }\href@noop {} {\bibfield
  {journal} {\bibinfo  {journal} {Applied Physics Letters}\ }\textbf {\bibinfo
  {volume} {92}},\ \bibinfo {pages} {221905} (\bibinfo {year}
  {2008})}\BibitemShut {NoStop}%
\bibitem [{\citenamefont {Sukhovich}\ \emph {et~al.}(2008)\citenamefont
  {Sukhovich}, \citenamefont {Jing},\ and\ \citenamefont
  {Page}}]{sukhovich2008negative}%
  \BibitemOpen
  \bibfield  {author} {\bibinfo {author} {\bibfnamefont {A.}~\bibnamefont
  {Sukhovich}}, \bibinfo {author} {\bibfnamefont {L.}~\bibnamefont {Jing}}, \
  and\ \bibinfo {author} {\bibfnamefont {J.}~\bibnamefont {Page}},\ }\href@noop
  {} {\bibfield  {journal} {\bibinfo  {journal} {Physical Review B}\ }\textbf
  {\bibinfo {volume} {77}},\ \bibinfo {pages} {014301} (\bibinfo {year}
  {2008})}\BibitemShut {NoStop}%
\bibitem [{\citenamefont {Lin}\ \emph {et~al.}(2009)\citenamefont {Lin},
  \citenamefont {Huang}, \citenamefont {Sun},\ and\ \citenamefont
  {Wu}}]{lin2009gradient}%
  \BibitemOpen
  \bibfield  {author} {\bibinfo {author} {\bibfnamefont {S.}~\bibnamefont
  {Lin}}, \bibinfo {author} {\bibfnamefont {T.}~\bibnamefont {Huang}}, \bibinfo
  {author} {\bibfnamefont {J.}~\bibnamefont {Sun}}, \ and\ \bibinfo {author}
  {\bibfnamefont {T.}~\bibnamefont {Wu}},\ }\href@noop {} {\bibfield  {journal}
  {\bibinfo  {journal} {Physical Review B}\ }\textbf {\bibinfo {volume} {79}},\
  \bibinfo {pages} {094302} (\bibinfo {year} {2009})}\BibitemShut {NoStop}%
\bibitem [{\citenamefont {Martinezsala}\ \emph {et~al.}(1995)\citenamefont
  {Martinezsala}, \citenamefont {Sancho}, \citenamefont {Sanchez},
  \citenamefont {G{\'o}mez}, \citenamefont {Llinares},\ and\ \citenamefont
  {Meseguer}}]{martinezsala1995sound}%
  \BibitemOpen
  \bibfield  {author} {\bibinfo {author} {\bibfnamefont {R.}~\bibnamefont
  {Martinezsala}}, \bibinfo {author} {\bibfnamefont {J.}~\bibnamefont
  {Sancho}}, \bibinfo {author} {\bibfnamefont {J.}~\bibnamefont {Sanchez}},
  \bibinfo {author} {\bibfnamefont {V.}~\bibnamefont {G{\'o}mez}}, \bibinfo
  {author} {\bibfnamefont {J.}~\bibnamefont {Llinares}}, \ and\ \bibinfo
  {author} {\bibfnamefont {F.}~\bibnamefont {Meseguer}},\ }\href@noop {}
  {\bibfield  {journal} {\bibinfo  {journal} {Nature}\ }\textbf {\bibinfo
  {volume} {378}},\ \bibinfo {pages} {241} (\bibinfo {year}
  {1995})}\BibitemShut {NoStop}%
\bibitem [{\citenamefont {Hussein}(2009)}]{hussein2009reduced}%
  \BibitemOpen
  \bibfield  {author} {\bibinfo {author} {\bibfnamefont {M.}~\bibnamefont
  {Hussein}},\ }\href@noop {} {\bibfield  {journal} {\bibinfo  {journal}
  {Proceedings of the Royal Society A: Mathematical, Physical and Engineering
  Science}\ }\textbf {\bibinfo {volume} {465}},\ \bibinfo {pages} {2825}
  (\bibinfo {year} {2009})}\BibitemShut {NoStop}%
\bibitem [{\citenamefont {Torquato}(2010)}]{torquato2010optimal}%
  \BibitemOpen
  \bibfield  {author} {\bibinfo {author} {\bibfnamefont {S.}~\bibnamefont
  {Torquato}},\ }\href@noop {} {\bibfield  {journal} {\bibinfo  {journal}
  {Annual Review of Materials Research}\ }\textbf {\bibinfo {volume} {40}},\
  \bibinfo {pages} {101} (\bibinfo {year} {2010})}\BibitemShut {NoStop}%
\bibitem [{\citenamefont {Sigmund}\ and\ \citenamefont
  {Jensen}(2003)}]{sigmund2003systematic}%
  \BibitemOpen
  \bibfield  {author} {\bibinfo {author} {\bibfnamefont {O.}~\bibnamefont
  {Sigmund}}\ and\ \bibinfo {author} {\bibfnamefont {J.~S.}\ \bibnamefont
  {Jensen}},\ }\href@noop {} {\bibfield  {journal} {\bibinfo  {journal}
  {Philosophical Transactions of the Royal Society of London. Series A:
  Mathematical, Physical and Engineering Sciences}\ }\textbf {\bibinfo {volume}
  {361}},\ \bibinfo {pages} {1001} (\bibinfo {year} {2003})}\BibitemShut
  {NoStop}%
\bibitem [{\citenamefont {Rupp}\ \emph {et~al.}(2007)\citenamefont {Rupp},
  \citenamefont {Evgrafov}, \citenamefont {Maute},\ and\ \citenamefont
  {Dunn}}]{rupp2007design}%
  \BibitemOpen
  \bibfield  {author} {\bibinfo {author} {\bibfnamefont {C.~J.}\ \bibnamefont
  {Rupp}}, \bibinfo {author} {\bibfnamefont {A.}~\bibnamefont {Evgrafov}},
  \bibinfo {author} {\bibfnamefont {K.}~\bibnamefont {Maute}}, \ and\ \bibinfo
  {author} {\bibfnamefont {M.~L.}\ \bibnamefont {Dunn}},\ }\href@noop {}
  {\bibfield  {journal} {\bibinfo  {journal} {Structural and Multidisciplinary
  Optimization}\ }\textbf {\bibinfo {volume} {34}},\ \bibinfo {pages} {111}
  (\bibinfo {year} {2007})}\BibitemShut {NoStop}%
\bibitem [{\citenamefont {Diaz}\ \emph {et~al.}(2005)\citenamefont {Diaz},
  \citenamefont {Haddow},\ and\ \citenamefont {Ma}}]{diaz2005design}%
  \BibitemOpen
  \bibfield  {author} {\bibinfo {author} {\bibfnamefont {A.}~\bibnamefont
  {Diaz}}, \bibinfo {author} {\bibfnamefont {A.}~\bibnamefont {Haddow}}, \ and\
  \bibinfo {author} {\bibfnamefont {L.}~\bibnamefont {Ma}},\ }\href@noop {}
  {\bibfield  {journal} {\bibinfo  {journal} {Structural and Multidisciplinary
  Optimization}\ }\textbf {\bibinfo {volume} {29}},\ \bibinfo {pages} {418}
  (\bibinfo {year} {2005})}\BibitemShut {NoStop}%
\bibitem [{\citenamefont {Halkj{\ae}r}\ \emph {et~al.}(2006)\citenamefont
  {Halkj{\ae}r}, \citenamefont {Sigmund},\ and\ \citenamefont
  {Jensen}}]{halkjaer2006maximizing}%
  \BibitemOpen
  \bibfield  {author} {\bibinfo {author} {\bibfnamefont {S.}~\bibnamefont
  {Halkj{\ae}r}}, \bibinfo {author} {\bibfnamefont {O.}~\bibnamefont
  {Sigmund}}, \ and\ \bibinfo {author} {\bibfnamefont {J.~S.}\ \bibnamefont
  {Jensen}},\ }\href@noop {} {\bibfield  {journal} {\bibinfo  {journal}
  {Structural and Multidisciplinary Optimization}\ }\textbf {\bibinfo {volume}
  {32}},\ \bibinfo {pages} {263} (\bibinfo {year} {2006})}\BibitemShut
  {NoStop}%
\bibitem [{\citenamefont {Sigmund}\ and\ \citenamefont
  {Hougaard}(2008)}]{sigmund2008geometric}%
  \BibitemOpen
  \bibfield  {author} {\bibinfo {author} {\bibfnamefont {O.}~\bibnamefont
  {Sigmund}}\ and\ \bibinfo {author} {\bibfnamefont {K.}~\bibnamefont
  {Hougaard}},\ }\href@noop {} {\bibfield  {journal} {\bibinfo  {journal}
  {Physical review letters}\ }\textbf {\bibinfo {volume} {100}},\ \bibinfo
  {pages} {153904} (\bibinfo {year} {2008})}\BibitemShut {NoStop}%
\bibitem [{\citenamefont {Bilal}\ and\ \citenamefont
  {Hussein}(2011)}]{bilal2011ultrawide}%
  \BibitemOpen
  \bibfield  {author} {\bibinfo {author} {\bibfnamefont {O.~R.}\ \bibnamefont
  {Bilal}}\ and\ \bibinfo {author} {\bibfnamefont {M.~I.}\ \bibnamefont
  {Hussein}},\ }\href@noop {} {\bibfield  {journal} {\bibinfo  {journal}
  {Physical Review E}\ }\textbf {\bibinfo {volume} {84}},\ \bibinfo {pages}
  {065701} (\bibinfo {year} {2011})}\BibitemShut {NoStop}%
\bibitem [{\citenamefont {Nemat-Nasser}(1972)}]{nemat1972harmonic}%
  \BibitemOpen
  \bibfield  {author} {\bibinfo {author} {\bibfnamefont {S.}~\bibnamefont
  {Nemat-Nasser}},\ }\href@noop {} {\bibfield  {journal} {\bibinfo  {journal}
  {Journal of Applied Mechanics}\ }\textbf {\bibinfo {volume} {39}},\ \bibinfo
  {pages} {850} (\bibinfo {year} {1972})}\BibitemShut {NoStop}%
\bibitem [{\citenamefont {Babuska}\ and\ \citenamefont
  {Osborn}(1978)}]{babuska1978numerical}%
  \BibitemOpen
  \bibfield  {author} {\bibinfo {author} {\bibfnamefont {I.}~\bibnamefont
  {Babuska}}\ and\ \bibinfo {author} {\bibfnamefont {J.}~\bibnamefont
  {Osborn}},\ }\href@noop {} {\bibfield  {journal} {\bibinfo  {journal} {Math.
  Comp}\ }\textbf {\bibinfo {volume} {32}},\ \bibinfo {pages} {991} (\bibinfo
  {year} {1978})}\BibitemShut {NoStop}%
\bibitem [{\citenamefont {Srivastava}\ and\ \citenamefont
  {Nemat-Nasser}(2013)}]{srivastava2013mixed}%
  \BibitemOpen
  \bibfield  {author} {\bibinfo {author} {\bibfnamefont {A.}~\bibnamefont
  {Srivastava}}\ and\ \bibinfo {author} {\bibfnamefont {S.}~\bibnamefont
  {Nemat-Nasser}},\ }\href@noop {} {\enquote {\bibinfo {title}
  {Mixed-variational formulation for phononic band-structure calculation of
  arbitrarily complex unit cells},}\ } (\bibinfo {year} {2013}),\ \bibinfo
  {note} {submitted}\BibitemShut {NoStop}%
\bibitem [{\citenamefont {Elsen}\ \emph {et~al.}(2007)\citenamefont {Elsen},
  \citenamefont {Vishal}, \citenamefont {Houston}, \citenamefont {Pande},
  \citenamefont {Hanrahan},\ and\ \citenamefont {Darve}}]{elsen2007n}%
  \BibitemOpen
  \bibfield  {author} {\bibinfo {author} {\bibfnamefont {E.}~\bibnamefont
  {Elsen}}, \bibinfo {author} {\bibfnamefont {V.}~\bibnamefont {Vishal}},
  \bibinfo {author} {\bibfnamefont {M.}~\bibnamefont {Houston}}, \bibinfo
  {author} {\bibfnamefont {V.}~\bibnamefont {Pande}}, \bibinfo {author}
  {\bibfnamefont {P.}~\bibnamefont {Hanrahan}}, \ and\ \bibinfo {author}
  {\bibfnamefont {E.}~\bibnamefont {Darve}},\ }\href@noop {} {\bibfield
  {journal} {\bibinfo  {journal} {arXiv preprint arXiv:0706.3060}\ } (\bibinfo
  {year} {2007})}\BibitemShut {NoStop}%
\bibitem [{\citenamefont {Stone}\ \emph {et~al.}(2007)\citenamefont {Stone},
  \citenamefont {Phillips}, \citenamefont {Freddolino}, \citenamefont {Hardy},
  \citenamefont {Trabuco},\ and\ \citenamefont
  {Schulten}}]{stone2007accelerating}%
  \BibitemOpen
  \bibfield  {author} {\bibinfo {author} {\bibfnamefont {J.~E.}\ \bibnamefont
  {Stone}}, \bibinfo {author} {\bibfnamefont {J.~C.}\ \bibnamefont {Phillips}},
  \bibinfo {author} {\bibfnamefont {P.~L.}\ \bibnamefont {Freddolino}},
  \bibinfo {author} {\bibfnamefont {D.~J.}\ \bibnamefont {Hardy}}, \bibinfo
  {author} {\bibfnamefont {L.~G.}\ \bibnamefont {Trabuco}}, \ and\ \bibinfo
  {author} {\bibfnamefont {K.}~\bibnamefont {Schulten}},\ }\href@noop {}
  {\bibfield  {journal} {\bibinfo  {journal} {Journal of computational
  chemistry}\ }\textbf {\bibinfo {volume} {28}},\ \bibinfo {pages} {2618}
  (\bibinfo {year} {2007})}\BibitemShut {NoStop}%
\bibitem [{\citenamefont {Owens}\ \emph {et~al.}(2008)\citenamefont {Owens},
  \citenamefont {Houston}, \citenamefont {Luebke}, \citenamefont {Green},
  \citenamefont {Stone},\ and\ \citenamefont {Phillips}}]{owens2008gpu}%
  \BibitemOpen
  \bibfield  {author} {\bibinfo {author} {\bibfnamefont {J.~D.}\ \bibnamefont
  {Owens}}, \bibinfo {author} {\bibfnamefont {M.}~\bibnamefont {Houston}},
  \bibinfo {author} {\bibfnamefont {D.}~\bibnamefont {Luebke}}, \bibinfo
  {author} {\bibfnamefont {S.}~\bibnamefont {Green}}, \bibinfo {author}
  {\bibfnamefont {J.~E.}\ \bibnamefont {Stone}}, \ and\ \bibinfo {author}
  {\bibfnamefont {J.~C.}\ \bibnamefont {Phillips}},\ }\href@noop {} {\bibfield
  {journal} {\bibinfo  {journal} {Proceedings of the IEEE}\ }\textbf {\bibinfo
  {volume} {96}},\ \bibinfo {pages} {879} (\bibinfo {year} {2008})}\BibitemShut
  {NoStop}%
\bibitem [{\citenamefont {Michalakes}\ and\ \citenamefont
  {Vachharajani}(2008)}]{michalakes2008gpu}%
  \BibitemOpen
  \bibfield  {author} {\bibinfo {author} {\bibfnamefont {J.}~\bibnamefont
  {Michalakes}}\ and\ \bibinfo {author} {\bibfnamefont {M.}~\bibnamefont
  {Vachharajani}},\ }\href@noop {} {\bibfield  {journal} {\bibinfo  {journal}
  {Parallel Processing Letters}\ }\textbf {\bibinfo {volume} {18}},\ \bibinfo
  {pages} {531} (\bibinfo {year} {2008})}\BibitemShut {NoStop}%
\bibitem [{\citenamefont {Humphrey}\ \emph {et~al.}(2006)\citenamefont
  {Humphrey}, \citenamefont {Price}, \citenamefont {Durbano}, \citenamefont
  {Kelmelis},\ and\ \citenamefont {Martin}}]{humphrey2006high}%
  \BibitemOpen
  \bibfield  {author} {\bibinfo {author} {\bibfnamefont {J.~R.}\ \bibnamefont
  {Humphrey}}, \bibinfo {author} {\bibfnamefont {D.~K.}\ \bibnamefont {Price}},
  \bibinfo {author} {\bibfnamefont {J.~P.}\ \bibnamefont {Durbano}}, \bibinfo
  {author} {\bibfnamefont {E.~J.}\ \bibnamefont {Kelmelis}}, \ and\ \bibinfo
  {author} {\bibfnamefont {R.~D.}\ \bibnamefont {Martin}},\ }in\ \href@noop {}
  {\emph {\bibinfo {booktitle} {Proceedings of the 10th WSEAS International
  Conference on APPLIED MATHEMATICS}}}\ (\bibinfo {year} {2006})\ pp.\ \bibinfo
  {pages} {547--550}\BibitemShut {NoStop}%
\bibitem [{\citenamefont {Nemat-Nasser}\ and\ \citenamefont
  {Srivastava}(2011)}]{nemat2011overall}%
  \BibitemOpen
  \bibfield  {author} {\bibinfo {author} {\bibfnamefont {S.}~\bibnamefont
  {Nemat-Nasser}}\ and\ \bibinfo {author} {\bibfnamefont {A.}~\bibnamefont
  {Srivastava}},\ }\href@noop {} {\bibfield  {journal} {\bibinfo  {journal}
  {Journal of the Mechanics and Physics of Solids}\ }\textbf {\bibinfo {volume}
  {59}},\ \bibinfo {pages} {1953} (\bibinfo {year} {2011})}\BibitemShut
  {NoStop}%
\bibitem [{\citenamefont {Srivastava}\ and\ \citenamefont
  {Nemat-Nasser}(2012)}]{srivastava2012overall}%
  \BibitemOpen
  \bibfield  {author} {\bibinfo {author} {\bibfnamefont {A.}~\bibnamefont
  {Srivastava}}\ and\ \bibinfo {author} {\bibfnamefont {S.}~\bibnamefont
  {Nemat-Nasser}},\ }\href@noop {} {\bibfield  {journal} {\bibinfo  {journal}
  {Proceedings of the Royal Society A: Mathematical, Physical and Engineering
  Science}\ }\textbf {\bibinfo {volume} {468}},\ \bibinfo {pages} {269}
  (\bibinfo {year} {2012})}\BibitemShut {NoStop}%
\bibitem [{\citenamefont {Shuvalov}\ \emph {et~al.}(2011)\citenamefont
  {Shuvalov}, \citenamefont {Kutsenko}, \citenamefont {Norris},\ and\
  \citenamefont {Poncelet}}]{shuvalov2011effective}%
  \BibitemOpen
  \bibfield  {author} {\bibinfo {author} {\bibfnamefont {A.}~\bibnamefont
  {Shuvalov}}, \bibinfo {author} {\bibfnamefont {A.}~\bibnamefont {Kutsenko}},
  \bibinfo {author} {\bibfnamefont {A.}~\bibnamefont {Norris}}, \ and\ \bibinfo
  {author} {\bibfnamefont {O.}~\bibnamefont {Poncelet}},\ }\href@noop {}
  {\bibfield  {journal} {\bibinfo  {journal} {Proceedings of the Royal Society
  A: Mathematical, Physical and Engineering Science}\ }\textbf {\bibinfo
  {volume} {467}},\ \bibinfo {pages} {1749} (\bibinfo {year}
  {2011})}\BibitemShut {NoStop}%
\bibitem [{\citenamefont {Willis}(2011)}]{willis2011effective}%
  \BibitemOpen
  \bibfield  {author} {\bibinfo {author} {\bibfnamefont {J.}~\bibnamefont
  {Willis}},\ }\href@noop {} {\bibfield  {journal} {\bibinfo  {journal}
  {Proceedings of the Royal Society A: Mathematical, Physical and Engineering
  Science}\ }\textbf {\bibinfo {volume} {467}},\ \bibinfo {pages} {1865}
  (\bibinfo {year} {2011})}\BibitemShut {NoStop}%
\bibitem [{\citenamefont {Willis}(2012)}]{willis2012construction}%
  \BibitemOpen
  \bibfield  {author} {\bibinfo {author} {\bibfnamefont {J.~R.}\ \bibnamefont
  {Willis}},\ }\href@noop {} {\bibfield  {journal} {\bibinfo  {journal}
  {Comptes Rendus M{\'e}canique}\ }\textbf {\bibinfo {volume} {340}},\ \bibinfo
  {pages} {181} (\bibinfo {year} {2012})}\BibitemShut {NoStop}%
\bibitem [{\citenamefont {Guest}\ and\ \citenamefont
  {Pr{\'e}vost}(2006)}]{guest2006optimizing}%
  \BibitemOpen
  \bibfield  {author} {\bibinfo {author} {\bibfnamefont {J.~K.}\ \bibnamefont
  {Guest}}\ and\ \bibinfo {author} {\bibfnamefont {J.~H.}\ \bibnamefont
  {Pr{\'e}vost}},\ }\href@noop {} {\bibfield  {journal} {\bibinfo  {journal}
  {International Journal of Solids and Structures}\ }\textbf {\bibinfo {volume}
  {43}},\ \bibinfo {pages} {7028} (\bibinfo {year} {2006})}\BibitemShut
  {NoStop}%
\bibitem [{\citenamefont {Guest}\ and\ \citenamefont
  {Pr{\'e}vost}(2007)}]{guest2007design}%
  \BibitemOpen
  \bibfield  {author} {\bibinfo {author} {\bibfnamefont {J.~K.}\ \bibnamefont
  {Guest}}\ and\ \bibinfo {author} {\bibfnamefont {J.~H.}\ \bibnamefont
  {Pr{\'e}vost}},\ }\href@noop {} {\bibfield  {journal} {\bibinfo  {journal}
  {Computer Methods in Applied Mechanics and Engineering}\ }\textbf {\bibinfo
  {volume} {196}},\ \bibinfo {pages} {1006} (\bibinfo {year}
  {2007})}\BibitemShut {NoStop}%
\bibitem [{\citenamefont {Minagawa}\ and\ \citenamefont
  {Nemat-Nasser}(1976)}]{minagawa1976harmonic}%
  \BibitemOpen
  \bibfield  {author} {\bibinfo {author} {\bibfnamefont {S.}~\bibnamefont
  {Minagawa}}\ and\ \bibinfo {author} {\bibfnamefont {S.}~\bibnamefont
  {Nemat-Nasser}},\ }\href@noop {} {\bibfield  {journal} {\bibinfo  {journal}
  {International Journal of Solids and Structures}\ }\textbf {\bibinfo {volume}
  {12}},\ \bibinfo {pages} {769} (\bibinfo {year} {1976})}\BibitemShut
  {NoStop}%
\bibitem [{\citenamefont {Geuzaine}\ and\ \citenamefont
  {Remacle}(2009)}]{geuzaine2009gmsh}%
  \BibitemOpen
  \bibfield  {author} {\bibinfo {author} {\bibfnamefont {C.}~\bibnamefont
  {Geuzaine}}\ and\ \bibinfo {author} {\bibfnamefont {J.-F.}\ \bibnamefont
  {Remacle}},\ }\href@noop {} {\bibfield  {journal} {\bibinfo  {journal}
  {International Journal for Numerical Methods in Engineering}\ }\textbf
  {\bibinfo {volume} {79}},\ \bibinfo {pages} {1309} (\bibinfo {year}
  {2009})}\BibitemShut {NoStop}%
\bibitem [{\citenamefont {Vasseur}\ \emph {et~al.}(2001)\citenamefont
  {Vasseur}, \citenamefont {Deymier}, \citenamefont {Chenni}, \citenamefont
  {Djafari-Rouhani}, \citenamefont {Dobrzynski},\ and\ \citenamefont
  {Prevost}}]{vasseur2001experimental}%
  \BibitemOpen
  \bibfield  {author} {\bibinfo {author} {\bibfnamefont {J.}~\bibnamefont
  {Vasseur}}, \bibinfo {author} {\bibfnamefont {P.}~\bibnamefont {Deymier}},
  \bibinfo {author} {\bibfnamefont {B.}~\bibnamefont {Chenni}}, \bibinfo
  {author} {\bibfnamefont {B.}~\bibnamefont {Djafari-Rouhani}}, \bibinfo
  {author} {\bibfnamefont {L.}~\bibnamefont {Dobrzynski}}, \ and\ \bibinfo
  {author} {\bibfnamefont {D.}~\bibnamefont {Prevost}},\ }\href@noop {}
  {\bibfield  {journal} {\bibinfo  {journal} {Physical Review Letters}\
  }\textbf {\bibinfo {volume} {86}},\ \bibinfo {pages} {3012} (\bibinfo {year}
  {2001})}\BibitemShut {NoStop}%
\end{thebibliography}


\end{document}